# Re-Thinking Schroedinger Boundary Conditions: the Simple Rotator is Not Simple

Arthur Davidson
ECE Dept., Carnegie Mellon University
Pittsburgh, PA 15213
artdav@ece.cmu.edu

ABSTRACT
A contradiction arises when applying standard boundary conditions to a simple quantum rotator with a single coordinate. New boundary conditions for the Schrödinger wave equation are proposed, that involve only gauge invariant quantities, and that can be applied beyond the immediate problem. The use of gauge invariant boundary conditions has important consequences for energy flow in or out of a simple rotator.

The well known boundary conditions for Schrödinger's equation are to use the wave function and its gradient to match smoothly across the boundary[1]. This prescription allows straightforward differentiation. Gauge invariant quantities like the probability density and the probability current density are automatically smooth and continuous across the boundary. For problems with continuous conservative potentials and in which the wave function goes to zero in all directions far from the boundary, everything works, and there are no loose ends.

A situation that is known to resist solution with these standard boundary conditions is the simple single-coordinate rotator. In this case the standard boundary conditions are that the wave function and its gradient should be periodic. The dilemma was very well characterized by Krieger and Iafrate[2]. Their work considers the closely related problem of periodicity in a 1D crystal and showed that periodic wave functions worked in one gauge, but not in another. Here these two salient choices of gauge are called the momentum gauge and the torque gauge. The momentum gauge works with the standard boundary conditions. Changing to the torque gauge, with an applied torque but no time dependent momentum, produces a wave function that that does not share the periodicity of the lattice. It is thus apparent that periodic wave function boundary conditions and their close counterpart in the simple rotator problem are not correct. The remedy is to look for boundary conditions that are themselves gauge invariant. The next two paragraphs give a brief overview of Table 1, which documents the argument. This is followed by a more detailed discussion, and conclusions.

The rows of Table 1, from top to bottom, present the argument for gauge invariant boundary conditions in the simple rotator problem. The two right hand columns show results for different parameters or equations in two gauges, designated the momentum gauge, and the torque gauge, for a 2D rigid rotator, which has a single coordinate, the angle $\theta$ running from $-\pi$ to $\pi$. Rows 2, 3, and 5, for canonical momentum, wave function, and Schrödinger equation, generally follow from Krieger and Iafrate [2]. These rows will be considered in more detail later. The action S in row 4, normalized by $\hbar$, is the overall phase factor of the wave function in each gauge choice. Row 6 is the critical row, which documents the standard boundary condition. This is the condition that the wave function itself, and its gradient, should be periodic for a rotator and for a 1 D crystal. Krieger and Iafrate [2] point out that this can work in the momentum gauge, where $u(\theta)$ is allowed to be periodic for an arbitrary choice of the real parameter $k/\hbar$. However, precisely because $u(\theta)$ is periodic and $k/\hbar$ is arbitrary this boundary condition cannot work in the torque gauge. The resolution of this essential contradiction is the purpose of this letter.

The boundary conditions in rows 7 and 8 are two equivalent formulations of gauge invariant boundary conditions that are sufficient for both the momentum and torque gauges. They are suggested by following line of reasoning: The Schrödinger equation written to explicitly show the dependence on amplitude and phase requires periodicity and smoothness in only 3 real quantities: the amplitude, the gradient of the amplitude, and the gradient of the phase (or gradient of the action S) [3]. The next several paragraphs and Figure 1 go into a more detailed description and discussion of Table 1.

The two salient gauges for the simple rotator problem are the momentum gauge and the torque gauge. In row 1 of Table 1, it is shown that there is no potential energy term for the Hamiltonian for the momentum gauge, while the torque gauge has a constant $T$ multiplying the coordinate, $\theta$. $T$ is the quantity of applied torque. The potential energy is a sloped line, and if $\theta$ goes from $-\pi$ to $\pi$ it is clear that the potential is necessarily discontinuous at $\pi$ and $-\pi$. It is the disappearance in the momentum gauge of this discontinuity in potential that allows Krieger and Iafrate to solve their problem using periodic wave function boundary conditions.

Row 2 in Table 1 is the usual prescription for canonical momentum: a linear potential energy term in the torque gauge can be replaced by a time-dependent momentum term in the canonical momentum in the momentum gauge. Row 3 in table 1 carries the consequences of a gauge change into the wave function phase. Note that the wave functions are not gauge invariant. Row 4 delineates the dependence of the action variables on the coordinate $\theta$ and time t. Note that the time–dependent momentum term $k(t)$ turns up explicitly in the action for the torque gauge.

Differences in the Schrödinger equation for the two gauges are detailed in row 5. The momentum term $k$ is explicit in the Hamiltonian for the momentum gauge, and there is also an ordinary differential equation to determine the dynamics of $k$. The torque gauge has the $k$ momentum term not in the Hamiltonian, but in the wave function $\Psi$ (as shown in row 3 and in Figure 1.)

Row 6 applies the standard boundary conditions in the two gauges. For the momentum gauge, the eigenfunction is proportional to $e^{i\lambda\theta}$, where the eigenvalue $\lambda$ is an integer. It works for the momentum gauge, because the value of $k/\hbar$, a real number, is removed from the wave function $u(\theta)$. Because this same arbitrary real number is part of the overall phase of the wave function in the torque gauge, applying the standard boundary conditions in this gauges in impossible. That is, the boundary condition requires both $\lambda$ and $k/\hbar$ to be integers, but the result from the momentum gauge is that $k/\hbar$ is not an integer. Thus the standard periodic wave function boundary condition leads to a contradiction when changing gauges.

**Figure 1** illustrates this dichotomy for eigenfunctions in the two gauges. Both gauges assume constant amplitude, and $\lambda=1$. In (a) on the left, the wave function phase $\lambda\theta$ goes through a full change of $2\pi$ as the coordinate $\theta$ goes from $-\pi$ to $\pi$. The starting and ending coordinates of the wave function are next to each other, so differentiating the wave function everywhere is straight forward. In (b), for the torque gauge, the ends of the wave function are separated by an angle proportional to the momentum parameter k (assumed to be some negative number for illustrative purposes). There is a jump in phase where there is a jump in potential. This phase jump violates the boundary condition that the wave function should be smooth across the boundary. If the boundary conditions are correct, then the only solutions allowed would be when $k/\hbar$ happens to be an integer, like $\lambda$. Since the momentum gauge is satisfied with real $k/\hbar$, it must be concluded that the periodic wave function boundary conditions are not valid in either gauge.

Rows 7 and 8 in Table 1 explore the use in boundary conditions of gauge invariant quantities related to the wave function. (Recall from row 3 that the wave function itself is not gauge invariant.) Using these quantities for boundary conditions goes beyond Krieger and Iafrate [2] because the problem should be solvable in any gauge, without changing the form of the boundary conditions. Row 7 uses the wave function amplitude $A$, $A_\theta$, and $S_\theta$ (the subscript denotes differentiation once by $\theta$). These 3 quantities are implied by the Schrödinger equation written explicitly for the real quantities A and S, in $\Psi$ in rows 3 and 4 of Table 1:

$$-\frac{\hbar^2}{2m_i}\frac{\partial^2 A}{\partial \theta^2} + \frac{1}{2m_i} A(\theta,t)\left(\frac{\partial S}{\partial \theta}\right)^2 - T\theta\, A = -A\frac{\partial S}{\partial t} \qquad (1)$$

$$2\frac{\partial A}{\partial \theta}\frac{\partial S}{\partial \theta} + A(\theta,t)\frac{\partial^2 S}{\partial \theta^2} = -2m_i \frac{\partial A}{\partial t} \qquad (2)$$

Equations 1 and 2 are the real and imaginary parts of Schrödinger's equation, written out for the real amplitude $A(\theta,t)$ and real action $S(\theta,t)$. The salient point is that this pair of equations is equivalent to the usual Schrödinger equation, and there are no terms for the action S lower than first order derivatives in the coordinate and time. The action itself does not appear. Therefore these equations will be satisfied at a boundary by continuity in $A$, $A_\theta$, and $S_\theta$ ($A$, $\nabla A$, and $\nabla S$ in higher dimensions). [4] Using these quantities in boundary conditions can be expected to have different results, in general, from the standard approach: the new boundary conditions involve continuity of 3 real quantities, whereas the standard boundary conditions involve continuity of 4 real quantities, since $\Psi$ and $\Psi_\theta$ are both complex.

Row 8 of Table 1 repeats the argument of row 7, but with three quantities that are both gauge invariant and more directly measureable: the probability density $\rho=A^2$, the gradient of that density, and the probability current density. The eigenfunctions in both the momentum gauge and torque gauge in rows 7 and 8 are consistent with each other: both $\lambda$ and $k/\hbar$ are real for these boundary conditions in any gauge. *What is remarkable* *about these new eigenfunctions is that they do not have quantized angular momentum—* that is, both $\lambda$ and $k/\hbar$ are any real numbers, no integers required. This implies that the simple rotator behaves just like a particle moving in one dimension along an infinite straight line: all real numbers are allowed as eigenvalues. With the gauge invariant boundary conditions discussed here, this unorthodox statement is true.

Where the rotator and the free particle differ is in the superposition of eigenstates. For the free particle, all eigenstates can be superposed, whereas for the rotator, only states with eigenvalues differing from $k/\hbar$ by an integer can be superposed. This is readily verified for periodic behavior of $\Psi^*\Psi$. The behavior of current density, action gradient, and probability density gradient are not given here but follow essentially the same logic. Let the torque T=0, so that true energy eigenfunctions exist. Then a general wave function and its complex conjugate can be expanded in terms of the momentum eigenfunctions found above:

$$\psi(\theta) = \sum_{j=-\infty}^{\infty} a_j e^{\frac{ik_j \theta}{\hbar}} \quad \text{and} \quad \psi(\theta)^* = \sum_{\ell=-\infty}^{\infty} a_\ell^* e^{\frac{-ik_\ell \theta}{\hbar}} . \tag{3}$$

Then $\Psi^*\Psi$ is given by:

$$\psi(\theta)^*\psi(\theta) = \sum_{\ell=-\infty}^{\infty}\sum_{j=-\infty}^{\infty} a_j a_\ell^* e^{\frac{i(k_j - k_\ell)\theta}{\hbar}} . \tag{4}$$

This means that $k_j$ and $k_\ell$ must differ by an integral multiple of $\hbar$ for $\Psi^*\Psi$ to be periodic:

$$k_j - k_\ell = integer \times \hbar . \tag{5}$$

It is always possible to pick an integer *n* such that $k_j = n\hbar + ko$ where $ko/\hbar$ is real. This fact together with Eq. 5 means that $k_\ell$ can be written as $k_\ell = m\hbar + ko$, where *m* is an integer.

Thus $\Psi^*\Psi$ will be periodic if $\psi(\theta) = \sum_{n=-\infty}^{\infty} a_n e^{i(n+\frac{k_0}{\hbar})\theta}$. The integer n here folds in the difference between $k_j$ and $k_\ell$ above. This is the same result as for the standard boundary conditions in the momentum gauge. But now it has been proven that it applies in all gauges, and that it results from applying the new gauge invariant boundary conditions to the superposition of eigenfunctions, rather than to each separate eigenfunction. For the rotator, then, the gauge invariant boundary conditions are applied twice: first to select the set of eigenfunctions with real eigenvalues, and then to select those superpositions that continue to satisfy the boundary conditions.

Now consider what happens to the eigenfunctions when $T \neq 0$. The form remains (in the torque gauge):

$$\psi\Big|_{T \neq 0} = A e^{i(n+\frac{k_0(t)}{\hbar})\theta} \tag{6}$$

Where A is a constant amplitude and now $k_0$ is time dependent. Substituting this form into the Schrödinger equation in the torque gauge, and collecting terms of similar symmetry, shows that the rate of change of $k_0$ is proportional to the torque T [5]. The function in Eq. 6 is a stationary state, but not strictly an eigenstate: The probability density $\rho$ is not changing, as would be true for an eigenfunction, but the eigenvalue is changing with time, reflecting the fact that an applied torque accelerates the rotator. Eq. 6 might be termed a "quasi-eigenfunction" Exactly the same band structure carries over to the finite torque case as for zero torque. The integer *n* in Eq. 6 is associated with the nth energy band of the system, while *k* changes continuously due to the torque. Once again, the momentum is not quantized, but the transitions from one band to another are.

Here is a summary of unorthodox results stemming from gauge invariant boundary conditions for the simple rotator:

1. For zero torque, all angular momentum eigenvalues are allowed, one at a time (no superposition). Like a free particle, the momentum is not quantized.

2. A constant torque accelerates the rotator and changes the total energy. This makes the rotator an open system, exchanging energy with a source in the environment. Under torque, the eigenfunctions become quasi-eigenfunctions, and remain stationary. However, the quasi-eigenvalues become time dependent.

3. Different eigenfunctions or quasi-eigenfunctions are not necessarily superposable, even though they each satisfy the gauge invariant boundary conditions.

4. Gauge invariant boundary conditions are applied to a superposition of (quasi-) eigenfunctions to determine if that superposition is allowed. The result is that all superposable (quasi-) eigenfunctions are separated in momentum by integral multiples of $\hbar$.

5. The rigid rotator and 1D periodic crystal are nearly the same mathematical problem, and use the same gauge invariant boundary conditions. Thus Bloch's theorem [6] applies to both. This means that Josephson junctions [7] and capacitors [5] have energy bands. These energy bands explain the observation of the coulomb blockade in small area capacitors and Josephson junctions, and may be important for analyzing Rabi oscillations in Josephson qubits [8, 9, 10].

The author acknowledges the support of the C2S2 Focus Center, one of five research centers funded under the Focus Center Research Program, a Semiconductor Research Corporation program.

Table 1 **Comparison of Gauge Choices for a Single-Coordinate Rotator**

| | **Momentum Gauge** | **Torque Gauge** |
|---|---|---|
| **1. Potential** | 0 | $T\theta$ |
| **2. Canonical momentum** | $-i\hbar\dfrac{\partial}{\partial x}+k,\ \dot{k}=T$ | $-i\hbar\dfrac{\partial}{\partial x}$ |
| **3. Wave function** | $u(\theta)=Ae^{i\phi/\hbar}$ | $\psi=u(\theta)e^{ik\theta/\hbar}$ <br> *If $u(\theta)$ is periodic, and $k/\hbar$ is arbitrary, then $\Psi(\theta)$ is not periodic* |
| **4. Action S** | $\phi(\theta,t)$ | $\phi(\theta,t)+k(t)\theta$ |
| **5. Schrödinger equation** | $\dfrac{1}{2m_i}(\dfrac{\hbar}{i}\dfrac{\partial}{\partial\theta}+k)^2 u(\theta,t)=i\hbar\dot{u}$ <br> $T=\dot{k}$ <br> *$k/\hbar$ = any real number* | $-\dfrac{\hbar^2}{2m_i}\dfrac{\partial^2}{\partial\theta^2}\psi(\theta,t)-T\theta\psi(\theta,t)=$ <br> $i\hbar\dot{\psi}$ |
| **Boundary Condition:** | | |
| **6. Periodic wave function:** | $u(\theta+2\pi)=u(\theta)$ <br> $u_\theta(\theta+2\pi)=u_\theta(\theta)$ <br> *Eigenfunction when T=0:* <br> $\sim e^{i\lambda\theta},\ \lambda$=integer | $\psi(\theta+2\pi)=\psi(\theta)$ <br> $\psi_\theta(\theta+2\pi)=\psi_\theta(\theta)$ <br> *Eigenfunction when T=0:* <br> $\sim e^{i(\lambda+k/\hbar)\theta},\ \lambda,\ k/\hbar$=integer |
| **7. $A,\ A_\theta,\ S_\theta$** | $A(\theta+2\pi)=A(\theta)$ <br> $A_\theta(\theta+2\pi)=A_\theta(\theta)$ <br> $S_\theta(\theta+2\pi)=S_\theta(\theta)$ <br> *Eigenfunction when T=0:* <br> $\sim e^{i\lambda\theta},\ \lambda$ = any real | Same as momentum gauge <br><br><br><br> *Eigenfunction when T=0:* <br> $\sim e^{i(\lambda+k/\hbar)\theta},\ \lambda,\ k/\hbar$ = any real |
| **8. $\rho=A^2,\ \rho_\theta,\ J(\theta)$** | $\rho(\theta+2\pi)=\rho(\theta)$ <br> $\rho_\theta(\theta+2\pi)=\rho_\theta(\theta)$ <br> $J_\theta(\theta+2\pi)=J_\theta(\theta)$ <br> *Eigenfunction when T=0:* <br> $\sim e^{i\lambda\theta},\ \lambda$ = any real | Same as momentum gauge <br><br><br><br> *Eigenfunction when T=0:* <br> $\sim e^{i(\lambda+k/\hbar)\theta},\ \lambda,\ k/\hbar$ = any real |

FIGURE

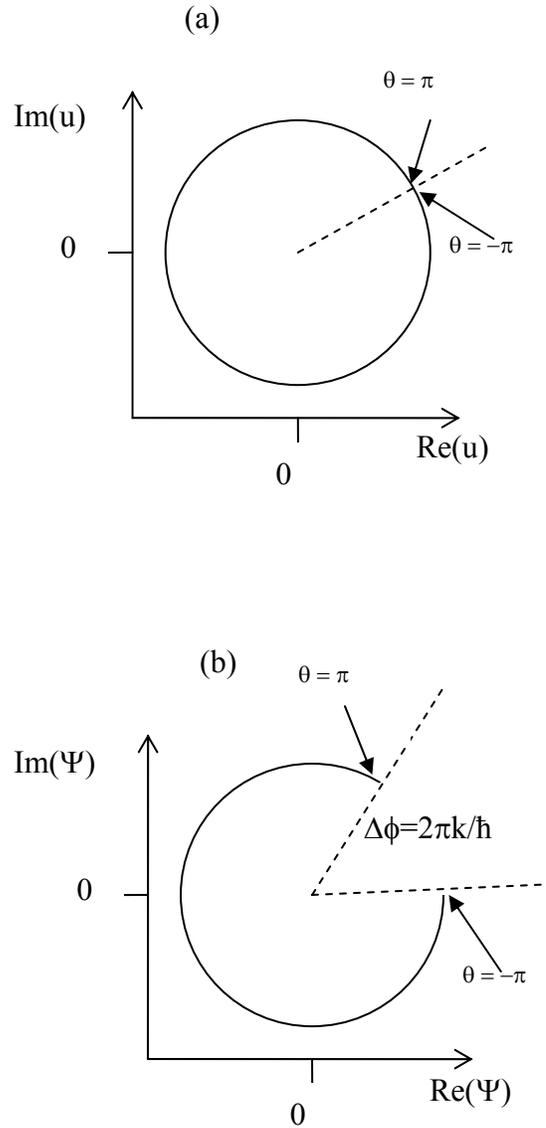

**Figure 1.** Plots of an eigenfunction for the simple rotator with zero torque. These are parametric plots in the complex plane, where the parameter is the coordinate $-\pi < \theta < \pi$. In (a) the eigenfunction $u(\theta)$ in the momentum gauge is closed and $2\pi$ periodic in $\theta$. In (b) the equivalent eigenfunction $\Psi(\theta)$ in the torque gauge is not periodic. With gauge invariant boundary conditions, both $u$ and $\Psi$ are allowed.

# REFERERENCES